\documentclass[%
 reprint,
  superscriptaddress,
 amsmath,amssymb,amsfonts,
 aps,
 prl,
 floatfix,
 longbibliography
]{revtex4-2}
\usepackage{comment}
\usepackage[utf8]{inputenc}
\usepackage[pdftex]{graphicx} \graphicspath{{}}
\usepackage{float} \usepackage{color}
\usepackage[pdftex,colorlinks=true]{hyperref}
\usepackage{subfigure}
\hypersetup{
  colorlinks=true,
  linkcolor=blue,
  citecolor = blue,
  urlcolor=blue,
}

\usepackage{mathtools}

\DeclarePairedDelimiter\ket{\lvert}{\rangle}
\DeclarePairedDelimiterX\braket[2]{\langle}{\rangle}{#1 \delimsize\vert #2}
\DeclarePairedDelimiterX\expval[3]{\langle}{\rangle}{#1 \delimsize\vert #2  \delimsize\vert #3}
\DeclarePairedDelimiterX\proj[2]{\delimsize\vert#1\rangle}{\langle#2\delimsize\vert}{ }
\usepackage{siunitx}

\usepackage{tikz}
\usetikzlibrary{arrows, arrows.meta}
\usepackage[T1]{fontenc}

\usepackage{tikz}
\definecolor{lime}{HTML}{A6CE39}
\DeclareRobustCommand{\orcidicon}{%
    \begin{tikzpicture}
    \draw[lime, fill=lime] (0,0) 
    circle [radius=0.13] 
    node[white] {{\fontfamily{qag}\selectfont\tiny ID}};
    \draw[white, fill=white] (-0.0625,0.095) 
    circle [radius=0.007];
    \end{tikzpicture}
    \hspace{-2mm}}

\newcommand{\orcidAD}{\href{https://orcid.org/0009-0000-0645-0772}{\orcidicon}}
\newcommand{\orcidLS}{\href{https://orcid.org/0000-0001-7652-9574}{\orcidicon}}

\usepackage{blindtext}
\begin{document}

\title{Anyon condensates of dipoles in triangular ladders}

\author{Arjo Dasgupta\orcidAD}
\email[]{arjo.dasgupta@itp.uni-hannover.de}
\affiliation{Institut f\"ur Theoretische Physik, Leibniz Universit\"at Hannover, Germany}

\author{Luis Santos\orcidLS}
\affiliation{Institut f\"ur Theoretische Physik, Leibniz Universit\"at Hannover, Germany}


\date{\today}

\begin{abstract}
Hard-core dipoles in triangular ladders are an excellent platform for the study of the interplay between frustration and long-range interactions, well described by a modified version of the celebrated $J_1$--$J_2$ model. Interestingly, as shown in [Phys. Rev. Lett. {\bf 109}, 227203 (2012)], such a model presents, for particular exactly-solvable conditions, a peculiar phase known as an anyon condensate. We show that balanced anyon condensates are robust against deviations from the exactly-solvable conditions, and discuss the requirements for dipolar orientation and ladder geometry, to realize anyon condensates of dipoles in triangular ladders, whose anyonic nature  may be easily revealed by time-of-flight measurements. Moreover, the ground-state physics in the vicinity of the exactly-solvable point is very rich, including a phase transition from anyon condensates into chiral superfluids, and self-bound Mott insulators, bond-order insulators, and chiral liquids.
\end{abstract}
\maketitle



Frustrated quantum magnets constitute a central topic in condensed-matter physics because competing interactions prevent the simultaneous minimization of all exchange couplings, giving rise to a wealth of unconventional phases~\cite{Balents2010}. Among them, the spin-$\frac12$ $J_1$-$J_2$ chain is a paradigmatic model of frustration induced by competing nearest- ($J_1$) and next-nearest-neighbor ($J_2$) exchange. Besides its exact dimerized ground state at the Majumdar-Ghosh point, $J_2=J_1/2$~\cite{Majumdar1969a,Majumdar1969b}, the model exhibits a rich phase diagram resulting from the interplay of frustration and quantum fluctuations~\cite{Okamoto1992,White1996}. Exchange anisotropy and magnetic fields further stabilize vector-chiral, multipolar, and spin-nematic phases~\cite{Vekua2007,Kecke2007,Hikihara2008,Sudan2009}.

A remarkable exception to this conventional picture was discovered by Batista and Somma, who showed that for fine-tuned exchange anisotropies the $J_1$-$J_2$ model admits exact ground states corresponding to condensates of one-dimensional anyons~\cite{Batista2012}. In these states the quasiparticles obey fractional exchange statistics characterized by a continuously tunable statistical phase. Unlike conventional Tomonaga-Luttinger liquids~(TLLs), including chiral TLLs~\cite{Nersesyan1998,Hikihara2008} and the non-chiral TLL$_2$ phase~\cite{Hikihara2010}, anyon condensates exhibit exponentially decaying single-particle correlations.

Ultracold atoms in optical lattices provide a versatile platform for the quantum simulation of frustrated magnetism, owing to the independent control of lattice geometry, tunneling, interactions, and synthetic gauge fields~\cite{Gross2017}. Recent experiments have realized frustrated magnetism in triangular optical lattices~\cite{Struck2011,Xu2023} and demonstrated long-range dipolar interactions using both magnetic atoms~\cite{DePaz2013,Baier2016,Su2023} and polar molecules~\cite{Yan2013,Christakis2023,Gregory2024,Carroll2025}. Combined with triangular-ladder geometries, dipolar interactions naturally realize frustrated $J_1$-$J_2$ and XXZ models and provide access to chiral and nematic phases in experimentally accessible regimes~\cite{Dasgupta2026}.

In this Letter, we show that properly oriented itinerant dipolar hard-core bosons in triangular ladders realize anyon-condensate phases. Whereas recent ultracold-atom experiments have demonstrated one-dimensional anyons through Floquet-engineered lattice dynamics~\cite{Kwan2024} and many-body anyonization in strongly interacting continuum gases~\cite{Dhar2025}, they probe few-body dynamics or emergent anyonic correlations rather than the condensation of anyonic quasiparticles. 



\begin{figure}[t!]
\centering
\includegraphics[width=0.8\columnwidth]{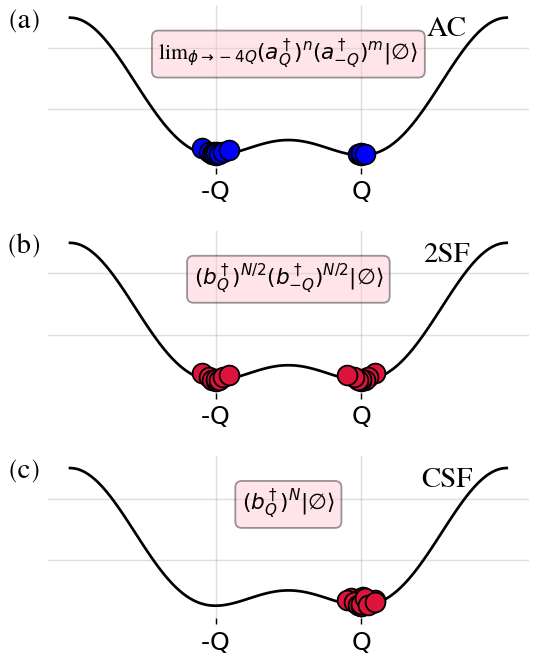}
\caption{Possible condensates in the frustrated $J_1-J_2$ model:  (a) Anyon condensate~(AC) state. (b) two-superfluid~(2SF) state. (c) Chiral superfluid~(CSF) state.  In the figure blue~(red) filled circles indicate anyons~(magnons).}
\label{fig:1}
\end{figure}


The realization of anyon condensates in ultracold quantum gases would therefore provide a qualitatively new route to explore the collective consequences of fractional statistics. We demonstrate that, although the extensive ground-state degeneracy of the exactly solvable model is lifted on deviating from the fine-tuned conditions of Ref.~\cite{Batista2012}, balanced anyon condensates survive over a broad parameter regime. Combining perturbation theory with DMRG calculations, we further uncover a rich phase diagram comprising transitions from anyon condensates to chiral superfluids, self-bound Mott insulators, bond-ordered insulators, and chiral liquids.



\begin{figure*}[t!]
\centering
\includegraphics[width=0.9\textwidth]{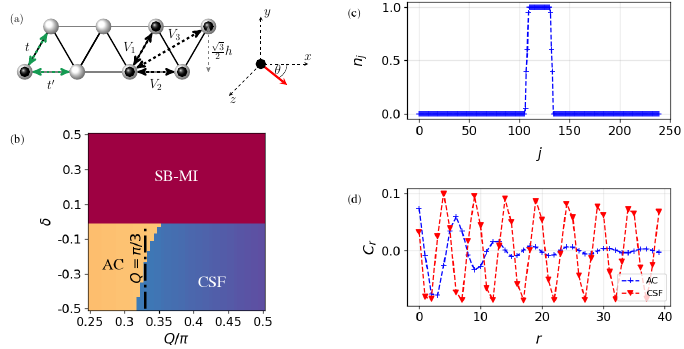}
\caption{(a) Scheme of the model of hard-core dipolar bosons on a triangular ladder of inter-leg separation $\frac{\sqrt 3}{2} h$, with dipoles polarized on the $xz$-plane at an angle $\theta$ with respect to the $x$-axis.
(b) Phase diagram of the $J_1$-$J_2$ model as a function of $Q$ and $\delta$ for a system of 240 sites with open boundary conditions at filling factor $\rho = 0.1$~($m=-0.4$). The phase boundaries are drawn from DMRG results with bond-dimension $\chi_{max}=800$ employing the TeNPy library \cite{Hauschild2018}. The dot-dashed line indicates the boundary between the AC and CSF phases as predicted by the perturbation theory. 
(c) Density profile of the self-bound Mott insulator at $Q=0.4\pi, \delta=0.2$.
(d) Comparison of single-particle correlations, $C_r \coloneqq \langle  \hat b^\dagger_j \hat b_{j+r} \rangle$, between CSF at $Q=0.4\pi$ and $\delta=-0.1$~(red triangles) and balanced-AC at $Q=0.28\pi$ and $\delta=-0.1$~(blue crosses).}
\label{fig:2}
\end{figure*}



\paragraph{Anyon condensates.--} Let us 
first briefly recall the concept of anyon condensates~\cite{Batista2012}. We consider a spin-$1/2$ 
$J_1-J_2$ XXZ Hamiltonian of the form:
\begin{equation}
\hat H_{J_1-J_2} = \sum_{j,\nu} J_\nu\left[
\frac{1}{2}\left( \hat S^+_j \hat S^-_{j+\nu} + \hat S^-_{j}\hat S^+_{j+\nu}\right ) + 
\Delta_\nu \hat S^z_j \hat S^z_{j+\nu} \right ],
\label{eq:HJ1J2}
\end{equation}
with $\hat S_j^{\pm,z}$ the spin operators associated to site $j$.
Geometric frustration demands antiferromagnetic coupling to the next-to-nearest neighbors, $J_2>0$. Moreover, if $J_2>|J_1|/4$, magnons present a two-well dispersion with non-equivalent minima at    quasi-momenta $\pm Q$, with 
$J_1/J_2 = -4\cos Q$~(see Figs.~\ref{fig:1}(a--c)).
As shown in Ref.~\cite{Batista2012}, for the particularly fine-tuned case in which 
$\Delta_\nu = \cos \nu Q$, 
the ground-state may be determined exactly~\cite{SM}. 
We introduce the  operator $\hat b_j^\dagger = \hat S_j^+$, which creates a magnon~(i.e. a spin $\uparrow$) at site $j$. We introduce as well the operators $\hat a_j^\dagger = e^{i\phi\sum_{l<j}(\hat S_l^z+1/2)} \hat S_j^+$, which create a magnon at site $j$, bound with a spin rotation of all spin spins to its left, $l<j$, by an angle $\phi$ around $z$. 
Whereas magnons are hard-core bosons, the operators 
$\hat a_j^\dagger$ fulfill anyonic commutation rules: 
$\hat a_j^\dag \hat a_{l>j}^\dag = e^{-i\phi} \hat a_l^\dag \hat a_j^\dag$, with the statistical angle $\phi=-4Q$. The ground-state of Hamiltonian~\eqref{eq:HJ1J2} is, under the above-mentioned conditions, given by the so-called anyon condensate (AC) states~(see Fig.~\ref{fig:1}~(a)):
\begin{eqnarray}
|\Psi_{n,m}(Q)\rangle &=& \frac{1}{\sqrt{{L \choose m+n}{m+n \choose m}}}\lim_{\phi\to -4Q} \left (\hat a_Q^\dag \right)^n \left (\hat a_{-Q}^\dag \right)^m 
|\varnothing\rangle \nonumber \\
&\propto& \!\!
\sum_{\substack{j_1<\dots < j_n \\ l_1<\dots <l_m}}
\prod_{q=1}^n e^{iQ j_q} \hat a_{j_q}^\dag  \prod_{r=1}^m \! e^{-iQ l_r} \hat a_{l_r}^\dag|\varnothing\rangle,
\end{eqnarray}
with $|\varnothing\rangle = |\downarrow, \dots, \downarrow\rangle$. These solutions are characterized by $n$~($m$) anyons populating the $Q$~($-Q$) dispersion minimum, see Fig.~\ref{fig:1}~(a). The overall number of anyons (or magnons)  is $n+m$. States with different anyon distributions between the two minima, i.e. different values of $n$ and $m$, are degenerate.  Anyon condensates should not be confused with a bi-condensate of magnons, 
$(\hat b_Q^\dagger)^{\otimes n} (\hat b_{-Q}^\dagger)^{\otimes m} |\varnothing\rangle$, see Fig.~\ref{fig:1}~(b). 
These bi-condensates would be a Luttinger liquid with conformal central charge $c=2$, characterized by a polynomial decay of correlation functions 
$C_r \equiv \langle \hat S_j^+ \hat S_{j+r}^-\rangle  = \langle
\hat b_j^\dagger \hat b_{j+r}\rangle$. This would correspond with the so-called two-superfluid, 2SF, phase of hard-core bosons~\cite{Mishra2015}, which in turn is equivalent to the TLL$_2$ phase of spin $J_1$-$J_2$ models~\cite{Hikihara2010}. In contrast, for anyon condensates $\langle \hat S_j^+ \hat S_{j+r}^-\rangle = \langle\hat a_j^\dagger   e^{i\phi\sum_{j\leq l < j+r} \hat a_l^\dagger \hat a_l}  \hat a_{j+r} \rangle $, where the exponent counts the number of anyons in between sites $j$ and $j+r$. Since in an AC state the number of anyons in between two sites has large fluctuations, this 
results in large phase fluctuations, leading to a characteristic exponential decay of correlations.
The states $\ket{\Psi_{N,0}(Q)}$ and $\ket{\Psi_{0,N}(Q)}$, with all particles condensed in either one of the two wells, see Fig.~\ref{fig:1}~(c), show no anyonic character. 
They are instead a magnon chiral superfluid~(CSF), with a
canted spin-spiral order~\cite{Batista2009}, characterised by long-range order in spin-spin correlations $C_r$ and a finite chirality $\kappa_j \coloneqq \left \langle  \left( \hat {\mathbf{S} }_j\times \hat{\mathbf{S}}_{j+1}\right)_z\right \rangle$. 


\paragraph{Realization using dipolar hard-core bosons.--} The exactly-solvable point of the $J_1-J_2$ model may be realized using hard-core itinerant dipolar bosons on a triangular optical ladder, see Fig.~\ref{fig:2}~(a). Assuming the lattice spacing along the legs~($x$ direction) as the length unit, 
the inter-leg separation~(along $y$) is parameterized as $h\sqrt{3}/2$, such that $h=1$ corresponds to 
equilateral triangular ladder. The dipoles are polarized along  the direction $\hat e_d = \cos \theta \hat e_x + \sin \theta \hat e_z$, see Fig.~\ref{fig:2}~(a). The bosons hop to the nearest neighbor with hopping strength $t=4\cos Q |t'|$ and to the next-nearest neighbor with $t'=-|t'|$. We choose $|t'|=1$ as our energy unit.
Dipoles separated by a vector  $(x,y)$ interact via the potential:
\begin{equation}
V(x,y) = \frac{V_0}{(x^2 + y^2)^{\frac{3}{2}}}\left(1-3\frac{x^2}{x^2+y^2} \cos ^2 \theta \right ).
\end{equation}
The dipolar lattice bosons are well described by the extended Bose-Hubbard model:
\begin{eqnarray}
\hat H_{B}&=& -\sum_{j}\Big( t\hat b^\dagger_j\hat b_{j+1}+t'\hat b^\dagger_j\hat b_{j+2} +\text{H.c.}\Big)\\
\nonumber
&+&\sum_j\sum_{r>0} V_r \hat n_j \hat n_{j+r},
\label{eq:EHB}
\end{eqnarray}
with $V_r=V(x_r, y_r)$, with 
$(x_r, y_r) = (\frac{r}{2},\frac{\sqrt 3}{2}h)$ for $r$ odd and $(x_r, y_r) = (\frac{r}{2},0)$ for $r$ even. 
Due to the hard-core nature of the bosons~($(\hat b_j^\dagger)^2=0$), $\hat H_{B}$ is equivalent, up to next-to-nearest neighbor interactions~($V_{r>2}=0$), to the Hamiltonian~\eqref{eq:HJ1J2} with $J_1=-2t$, $J_2=-2t'$, $\Delta_1=-V_1/2t$, and $\Delta_2=-V_2/2t'$. The exactly-solvable case, $\Delta_\nu = \cos(\nu Q)$, is fulfilled if $\theta=\theta_c$, with $\cos^2\theta_c=F(q,h)$, and $V_0=V_{0;c}=\frac{2\cos(2Q)}{1-3F(Q,h)}$, with 
\begin{eqnarray}
F(Q,h) &\equiv& \frac{1}{3}\left [\frac{4\cos^2 Q \left(\frac{1+3h^2}{4}\right)^\frac{3}{2} + \cos 2Q}{4\cos^2 Q \left(\frac{1+3h^2}{4}\right)^\frac{3}{2} +\frac{\cos 2Q}{1+3h^2}}\right ]
\label{eqn_theta}
\end{eqnarray}
The exactly-solvable case is only reachable if 
$|F(Q,h)|\leq 1$, which excludes a small region of $(Q,h)$ values, see~\cite{SM}. Note also that for some $(Q,h)$ values, $V_{0,c}<0$~(anti-dipolar configuration), which may be achieved 
by fast rotation of the applied magnetic field for the case of magnetic atoms~\cite{Giovanazzi2002, Tang2018}, or microwave dressing in the case of polar molecules~\cite{Karman2025}. For more details see~\cite{SM}.


\paragraph{Robustness of the anyon condensate against deviations from the exactly solvable point.--} The exactly-solvable point demands a fine tuning of $\Delta_{1,2}$. An interesting and relevant question concerns the 
robustness of the anyon condensate against deviations from the exactly-solvable point.
Let us consider in particular that the dipole strength deviates from $V_{0,c}$ into $V_0 = V_{0,c} (1+\delta)$, which results in $\Delta_\nu = \cos(\nu Q)(1+\delta)$. 
Perturbation theory for $\delta<0$ shows that for a low magnon filling, the energy 
of the anyon condensate acquires a correction that depends on the population $n$ and $m$ of the two dispersion minima as:
\begin{equation}
\Delta E_{n,m} = J_2 (1-4\cos^2 Q)\, |\delta| \frac{nm}{L^2},
\end{equation}
with $L$ the number of sites.
Since $J_2>0$, for $Q>\pi/3$~($t<2$), the energy 
is minimized for $n=0$ or $m=0$. Hence, for $Q>\pi/3$, the anyon condensate is not robust, and a slight deviation $\delta<0$ from the exactly-solvable case, results in a transition into a chiral superfluid. In contrast, for $Q<\pi/3$~($t>2$), the energy is minimized for $n=m$. Hence, for $Q<\pi/3$, the massive degeneracy of the anyon condensate solutions is lifted, but a balanced anyon condensate is expected to survive.

Numerical DMRG calculations of the model including interactions up to the next-to-nearest neighbour, for a lattice of $L=240$ sites with open boundary conditions and lattice filling $\rho=0.1$~(which corresponds to a spin model with magnetization $m=\langle \hat S_z \rangle = -0.4$) confirm that for $\delta<0$ there is a transition between a chiral superfluid~(CSF) and a balanced anyon condensate~(AC), see Fig.~\ref{fig:2}~(b). The CSF is characterized by finite long-range chirality-chirality correlations $\kappa^2 \coloneqq \lim_{|i-j| \rightarrow \infty }\langle \hat \kappa_i\hat \kappa_j\rangle$, with $\hat\kappa_j = \frac{1}{2i} (\hat b^\dagger_{j+1} \hat b_j - \hat b^\dagger_{j} \hat b_{j+1})$ the current operator, and by polynomically-decaying $C_r$. In contrast, the AC phase exhibits vanishing chirality~($\kappa^2=0$), and an exponentially-decaying $C_r \sim e^{-r/\xi}$~(see Fig.~\ref{fig:2}~(d)), with $\xi\propto 1/\rho$. As expected from the perturbative calculation, the transition takes place in the vicinity of $Q=\pi/3$.

For $\delta>0$, attractive inter-site interactions induce a gas-to-solid transition~\cite{Morera2023}, resulting in a self-bound Mott insulator (SB-MI), see Fig.~\ref{fig:2}~(c), in which all particles gather in a finite region with unit filling.



\begin{figure*}[t!]
\centering
\includegraphics[width=0.95\textwidth]{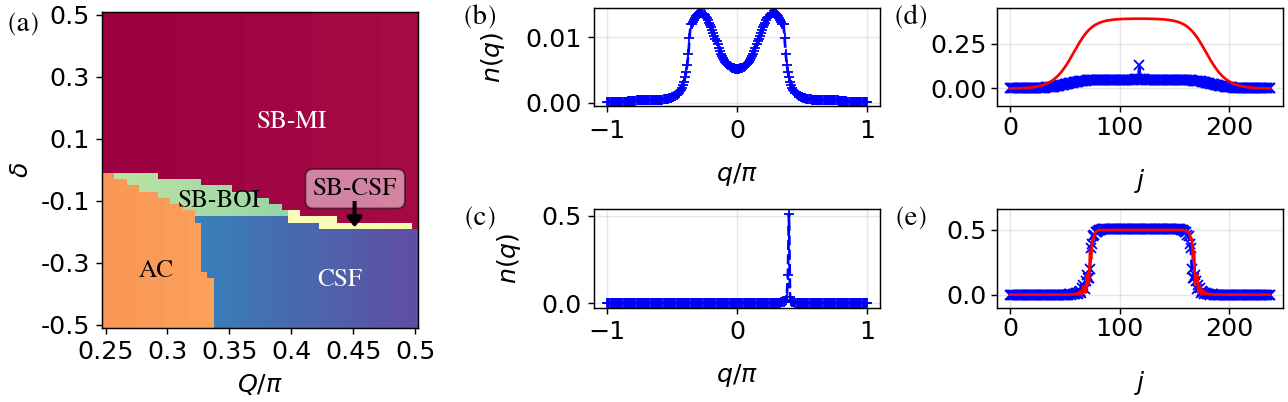}
\caption{(a) Phase diagram of hard-core dipolar bosons in a triangular ladder with $h=0.2$, as a function of $Q$ and $\delta$ for a system of $L=240$ sites with open boundary conditions at a filling $\rho=0.2$~(
$m = -0.3$). The phase boundaries are drawn from DMRG results with bond-dimension $\chi_{max}=800$ employing the TeNPy library \cite{Hauschild2018}. (b) Quasi-momentum distribution $n(q)$ for the balanced-AC ground-state at $Q=0.28\pi$, $\delta=-0.1$. (c) Quasi-momentum distribution for the CSF ground-state at $Q=0.4\pi$, $\delta=-0.2$. (d) Density profile~(red line) and chirality-chirality correlations~(blue crosses) with respect to the central site, $\langle \hat \kappa_{L/2}\hat \kappa_j\rangle$,~(blue) in the self-bound chiral superfluid~(SB-CSF) ground-state at $Q=0.4\pi$ and $\delta=-0.16$. (e) Density profile~(red line) and bond-order in absolute value, $|O_{BO}|$,~(blue crosses) in the self-bound bond-order insulator~(SB-BOI) ground-state at  $Q=\frac{\pi}{3}$ and $\delta=-0.06$.}
\label{fig:3}
\end{figure*}


\paragraph{The role of the dipolar tail.--} In the previous discussion, we considered a cut-off dipolar interaction. The inherent tail of the dipolar interaction, $V_{r>2}$ may however play an interesting role. As shown in Fig.~\ref{fig:3}~(a) for $\rho=0.2$~($m=-0.3$), the transition between CSF and the balanced AC is preserved when considering the actual dipolar interaction, hence confirming that balanced anyon condensates could be observed in dipolar lattice gases even under conditions that significantly deviate from the exactly solvable point. 
This transition may be easily investigated experimentally by monitoring the momentum distribution in time-of-flight experiments. Whereas the quasi-momentum distribution $n(q) \coloneqq \frac{1}{L^2}\sum_{jk} e^{iq(j-k)} \langle \hat b^\dagger_j \hat b_k\rangle$~of the CSF phase is characterized by a narrow peak at $Q$ or $-Q$, the balanced AC presents two symmetric broad peaks~(note that a bi-condensate would result in two equal narrow peaks at $+Q$ and $-Q$), see Figs.~\ref{fig:3}~(b) and (c).

The gas-to-solid transition is shifted below $\delta=0$. This occurs because $V_3$ and $V_4$ become attractive for $\frac{\pi}{4}<Q<\frac{\pi}{2}$, hence strengthening the tendency of self-binding. Interestingly, the attractive $V_3$ leads to the appearance of two novel self-bound phases for $\delta<0$. 
For $Q \gtrsim 0.38\pi$, a self-bound chiral superfluid (SB-CSF) or 
"chiral liquid" phase appears in between the CSF and SB-MI phases . The SB-CSF is characterized by a flat-top profile at a density $\rho<1$, and exhibits finite chirality correlations, see Fig.~\ref{fig:3}~(d). 
Moreover, for smaller values of $Q$, we find a self-bound bond-order insulator (SB-BOI) phase with a 
flat-top density profile at fixed density $\rho=1/2$. This phase is characterized by bond-order correlations $O_{BO} = \langle \hat b^\dagger_j \hat b_{j+1} - \hat b^\dagger_{j+1}\hat b_{j+2}\rangle \sim \pm 1/2$, oscillating from site to site, which indicate the separation of the state into a product of resonating dimers, every two links, $\ket{\Psi_0} = \bigotimes_{n} \frac{\ket{01}_{2n-1,2n}+\ket{10}_{2n-1,2n}}{\sqrt 2}$, see Fig.~\ref{fig:3}~(e). The appearance of this phase in the vicinity of $Q=\frac{\pi}{3}$ may be understood from the properties of the anisotropic Majumdar-Ghosh~(MG) model~\cite{Majumdar1970, Pal2021}. Note that for $Q=\pi/3$, and after rotating spins at even sites around the $z$ axis~($\hat S_{i\in \mathrm{even}}^{x,y}\to -\hat S_{i\in \mathrm{even}}^{x,y}$), the $J_1-J_2$ model is characterized by $J_2=J_1/2$~(MG ratio), and $\Delta_{1,2}=\Delta=-\frac{1}{2}(1+\delta)$. In our model, for $\delta<0$ attractive interactions beyond next-to-nearest neighbors induce self-binding with $\rho=1/2$ within the droplet, for sufficiently small $|\delta|$. For $\rho=1/2$~(zero magnetization in the spin model), the ground-state of the MG model is a dimer state for $\Delta>-1/2$~\cite{Pal2021}, i.e. a BOI for $\delta<0$, as we observe in the vicinity of $\delta=0$ and $Q=\pi/3$.


\paragraph{Conclusions.--} Itinerant dipolar gases in triangular optical ladders constitute a suitable platform for the realization of anyon condensates, whose nature may be easily revealed from the momentum distribution obtained in time-of-flight experiments. Although 
highly-degenerate anyon condensates were previously discussed only for peculiar fine-tuned conditions~\cite{Batista2012}, we have shown that, although the degeneracy is lost, balanced anyon condensates survive in a finite parameter regime. Furthermore, the vicinity of the exactly solvable point exhibits a very rich ground-state physics, characterized by a phase transition between a balanced anyon-condensate and a chiral superfluid, and the presence of 
self-bound solutions~(Mott insulator, bond-order insulator, and chiral liquid).  
The realization of anyon condensates using dipoles in optical arrays would provide a novel route to explore fractional statistics in ultracold gases.


\acknowledgments
We thank T. Posske for enlightening discussions. We acknowledge the support of the Deutsche Forschungsgemeinschaft (DFG, German Research Foundation) -- Project-ID 274200144 -- SFB 1227 DQ-mat within the project A04, and under Germany's Excellence Strategy -- EXC-2123 Quantum-Frontiers -- 390837967.


\bibliography{anyon_bib}

@article{Mishra2015,
  title = {Polar molecules in frustrated triangular ladders},
  author = {Mishra, Tapan and Greschner, Sebastian and Santos, Luis},
  journal = {Phys. Rev. A},
  volume = {91},
  issue = {4},
  pages = {043614},
  numpages = {8},
  year = {2015},
  month = {Apr},
  publisher = {American Physical Society},
  doi = {10.1103/PhysRevA.91.043614},
  url = {https://link.aps.org/doi/10.1103/PhysRevA.91.043614}
}

@article{Dhar2025,
	Author = {Dhar, Sudipta and Wang, Botao and Horvath, Milena and Vashisht, Amit and Zeng, Yi and Zvonarev, Mikhail B. and Goldman, Nathan and Guo, Yanliang and Landini, Manuele and N{\"a}gerl, Hanns-Christoph},
	Da = {2025/06/01},
	Doi = {10.1038/s41586-025-09016-9},
	Id = {Dhar2025},
	Isbn = {1476-4687},
	Journal = {Nature},
	Number = {8066},
	Pages = {53--57},
	Title = {Observing anyonization of bosons in a quantum gas},
	Ty = {JOUR},
	Url = {https://doi.org/10.1038/s41586-025-09016-9},
	Volume = {642},
	Year = {2025}}

@article{
Kwan2024,
author = {Joyce Kwan  and Perrin Segura  and Yanfei Li  and Sooshin Kim  and Alexey V. Gorshkov  and Andre Eckardt  and Brice Bakkali-Hassani  and Markus Greiner },
title = {Realization of one-dimensional anyons with arbitrary statistical phase},
journal = {Science},
volume = {386},
number = {6725},
pages = {1055-1060},
year = {2024},
doi = {10.1126/science.adi3252},
URL = {https://www.science.org/doi/abs/10.1126/science.adi3252}
}

@article{DePaz2013,
  title = {Nonequilibrium Quantum Magnetism in a Dipolar Lattice Gas},
  author = {de Paz, A. and Sharma, A. and Chotia, A. and Mar\'echal, E. and Huckans, J. H. and Pedri, P. and Santos, L. and Gorceix, O. and Vernac, L. and Laburthe-Tolra, B.},
  journal = {Phys. Rev. Lett.},
  volume = {111},
  issue = {18},
  pages = {185305},
  numpages = {5},
  year = {2013},
  month = {Oct},
  publisher = {American Physical Society},
  doi = {10.1103/PhysRevLett.111.185305},
  url = {https://link.aps.org/doi/10.1103/PhysRevLett.111.185305}
}

@article{Carroll2025,
author = {Annette N. Carroll  and Henrik Hirzler  and Calder Miller  and David Wellnitz  and Sean R. Muleady  and Junyu Lin  and Krzysztof P. Zamarski  and Reuben R. W. Wang  and John L. Bohn  and Ana Maria Rey  and Jun Ye },
title = {Observation of generalized <i>t-J</i> spin dynamics with tunable dipolar interactions},
journal = {Science},
volume = {388},
number = {6745},
pages = {381-386},
year = {2025},
doi = {10.1126/science.adq0911},
URL = {https://www.science.org/doi/abs/10.1126/science.adq0911}}

@article{Gregory2024,
	Author = {Gregory, Philip D. and Fernley, Luke M. and Tao, Albert Li and Bromley, Sarah L. and Stepp, Jonathan and Zhang, Zewen and Kotochigova, Svetlana and Hazzard, Kaden R. A. and Cornish, Simon L.},
	Da = {2024/03/01},
	Doi = {10.1038/s41567-023-02328-5},
	Id = {Gregory2024},
	Isbn = {1745-2481},
	Journal = {Nature Physics},
	Number = {3},
	Pages = {415--421},
	Title = {Second-scale rotational coherence and dipolar interactions in a gas of ultracold polar molecules},
	Ty = {JOUR},
	Url = {https://doi.org/10.1038/s41567-023-02328-5},
	Volume = {20},
	Year = {2024}}

@article{Christakis2023,
	Author = {Christakis, Lysander and Rosenberg, Jason S. and Raj, Ravin and Chi, Sungjae and Morningstar, Alan and Huse, David A. and Yan, Zoe Z. and Bakr, Waseem S.},
	Da = {2023/02/01},
	Doi = {10.1038/s41586-022-05558-4},
	Id = {Christakis2023},
	Isbn = {1476-4687},
	Journal = {Nature},
	Number = {7946},
	Pages = {64--69},
	Title = {Probing site-resolved correlations in a spin system of ultracold molecules},
	Ty = {JOUR},
	Url = {https://doi.org/10.1038/s41586-022-05558-4},
	Volume = {614},
	Year = {2023}}

@article{Yan2013,
	Author = {Yan, Bo and Moses, Steven A. and Gadway, Bryce and Covey, Jacob P. and Hazzard, Kaden R. A. and Rey, Ana Maria and Jin, Deborah S. and Ye, Jun},
	Date = {2013/09/01},
	Doi = {10.1038/nature12483},
	Id = {Yan2013},
	Isbn = {1476-4687},
	Journal = {Nature},
	Number = {7468},
	Pages = {521--525},
	Title = {Observation of dipolar spin-exchange interactions with lattice-confined polar molecules},
	Ty = {JOUR},
	Url = {https://doi.org/10.1038/nature12483},
	Volume = {501},
	Year = {2013}}

@article{Xu2023,
	Author = {Xu, Muqing and Kendrick, Lev Haldar and Kale, Anant and Gang, Youqi and Ji, Geoffrey and Scalettar, Richard T. and Lebrat, Martin and Greiner, Markus},
	Da = {2023/08/01},
	Doi = {10.1038/s41586-023-06280-5},
	Id = {Xu2023},
	Isbn = {1476-4687},
	Journal = {Nature},
	Number = {7976},
	Pages = {971--976},
	Title = {Frustration- and doping-induced magnetism in a Fermi--Hubbard simulator},
	Ty = {JOUR},
	Url = {https://doi.org/10.1038/s41586-023-06280-5},
	Volume = {620},
	Year = {2023}}

@article{Struck2011,
author = {J. Struck  and C. Ölschläger  and R. Le Targat  and P. Soltan-Panahi  and A. Eckardt  and M. Lewenstein  and P. Windpassinger  and K. Sengstock },
title = {Quantum Simulation of Frustrated Classical Magnetism in Triangular Optical Lattices},
journal = {Science},
volume = {333},
number = {6045},
pages = {996-999},
year = {2011},
doi = {10.1126/science.1207239},
URL = {https://www.science.org/doi/abs/10.1126/science.1207239}
}

@article{Gross2017,
author = {Christian Gross  and Immanuel Bloch },
title = {Quantum simulations with ultracold atoms in optical lattices},
journal = {Science},
volume = {357},
number = {6355},
pages = {995-1001},
year = {2017},
doi = {10.1126/science.aal3837},
URL = {https://www.science.org/doi/abs/10.1126/science.aal3837}}

@article{Nersesyan1998,
  title = {Incommensurate Spin Correlations in Spin- $1/2$ Frustrated Two-Leg Heisenberg Ladders},
  author = {Nersesyan, Alexander A. and Gogolin, Alexander O. and E\ss{}ler, Fabian H. L.},
  journal = {Phys. Rev. Lett.},
  volume = {81},
  issue = {4},
  pages = {910--913},
  numpages = {0},
  year = {1998},
  month = {Jul},
  publisher = {American Physical Society},
  doi = {10.1103/PhysRevLett.81.910},
  url = {https://link.aps.org/doi/10.1103/PhysRevLett.81.910}
}

@article{Sudan2009,
  title = {Emergent multipolar spin correlations in a fluctuating spiral: The frustrated ferromagnetic spin-$\frac{1}{2}$ Heisenberg chain in a magnetic field},
  author = {Sudan, Julien and L\"uscher, Andreas and L\"auchli, Andreas M.},
  journal = {Phys. Rev. B},
  volume = {80},
  issue = {14},
  pages = {140402(R)},
  numpages = {4},
  year = {2009},
  month = {Oct},
  publisher = {American Physical Society},
  doi = {10.1103/PhysRevB.80.140402},
  url = {https://link.aps.org/doi/10.1103/PhysRevB.80.140402}
}

@article{Hikihara2008,
  title = {Vector chiral and multipolar orders in the spin-$\frac{1}{2}$ frustrated ferromagnetic chain in magnetic field},
  author = {Hikihara, Toshiya and Kecke, Lars and Momoi, Tsutomu and Furusaki, Akira},
  journal = {Phys. Rev. B},
  volume = {78},
  issue = {14},
  pages = {144404},
  numpages = {19},
  year = {2008},
  month = {Oct},
  publisher = {American Physical Society},
  doi = {10.1103/PhysRevB.78.144404},
  url = {https://link.aps.org/doi/10.1103/PhysRevB.78.144404}
}

@article{Kecke2007,
  title = {Multimagnon bound states in the frustrated ferromagnetic one-dimensional chain},
  author = {Kecke, Lars and Momoi, Tsutomu and Furusaki, Akira},
  journal = {Phys. Rev. B},
  volume = {76},
  issue = {6},
  pages = {060407(R)},
  numpages = {4},
  year = {2007},
  month = {Aug},
  publisher = {American Physical Society},
  doi = {10.1103/PhysRevB.76.060407},
  url = {https://link.aps.org/doi/10.1103/PhysRevB.76.060407}
}

@article{Vekua2007,
  title = {Correlation functions and excitation spectrum of the frustrated ferromagnetic spin-$\frac{1}{2}$ chain in an external magnetic field},
  author = {Vekua, T. and Honecker, A. and Mikeska, H.-J. and Heidrich-Meisner, F.},
  journal = {Phys. Rev. B},
  volume = {76},
  issue = {17},
  pages = {174420},
  numpages = {6},
  year = {2007},
  month = {Nov},
  publisher = {American Physical Society},
  doi = {10.1103/PhysRevB.76.174420},
  url = {https://link.aps.org/doi/10.1103/PhysRevB.76.174420}
}

@article{Okamoto1992,
title = {Fluid-dimer critical point in S = 12 antiferromagnetic Heisenberg chain with next nearest neighbor interactions},
journal = {Physics Letters A},
volume = {169},
number = {6},
pages = {433-437},
year = {1992},
issn = {0375-9601},
doi = {https://doi.org/10.1016/0375-9601(92)90823-5},
url = {https://www.sciencedirect.com/science/article/pii/0375960192908235},
author = {Kiyomi Okamoto and Kiyohide Nomura}
}

@article{White1996,
  title = {Dimerization and incommensurate spiral spin correlations in the zigzag spin chain: Analogies to the Kondo lattice},
  author = {White, Steven R. and Affleck, Ian},
  journal = {Phys. Rev. B},
  volume = {54},
  issue = {14},
  pages = {9862--9869},
  numpages = {0},
  year = {1996},
  month = {Oct},
  publisher = {American Physical Society},
  doi = {10.1103/PhysRevB.54.9862},
  url = {https://link.aps.org/doi/10.1103/PhysRevB.54.9862}
}

@article{Majumdar1969b,
    author = {Majumdar, Chanchal K. and Ghosh, Dipan K.},
    title = {On Next‐Nearest‐Neighbor Interaction in Linear Chain. II},
    journal = {Journal of Mathematical Physics},
    volume = {10},
    number = {8},
    pages = {1399-1402},
    year = {1969},
    month = {08},
    issn = {0022-2488},
    doi = {10.1063/1.1664979},
    url = {https://doi.org/10.1063/1.1664979},
    eprint = {https://pubs.aip.org/aip/jmp/article-pdf/10/8/1399/19219508/1399\_1\_online.pdf},
}

@article{Majumdar1969a,
    author = {Majumdar, Chanchal K. and Ghosh, Dipan K.},
    title = {On Next‐Nearest‐Neighbor Interaction in Linear Chain. I},
    journal = {Journal of Mathematical Physics},
    volume = {10},
    number = {8},
    pages = {1388-1398},
    year = {1969},
    month = {08},
    issn = {0022-2488},
    doi = {10.1063/1.1664978},
    url = {https://doi.org/10.1063/1.1664978},
    eprint = {https://pubs.aip.org/aip/jmp/article-pdf/10/8/1388/19219090/1388\_1\_online.pdf},
}

@article{Pal2021,
  title = {Colorful points in the XY regime of XXZ quantum magnets},
  author = {Pal, Santanu and Sharma, Prakash and Changlani, Hitesh J. and Pujari, Sumiran},
  journal = {Phys. Rev. B},
  volume = {103},
  issue = {14},
  pages = {144414},
  numpages = {20},
  year = {2021},
  month = {Apr},
  publisher = {American Physical Society},
  doi = {10.1103/PhysRevB.103.144414},
  url = {https://link.aps.org/doi/10.1103/PhysRevB.103.144414}
}

@article{Karman2025,
  title = {Double Microwave Shielding},
  author = {Karman, Tijs and Bigagli, Niccol\`o and Yuan, Weijun and Zhang, Siwei and Stevenson, Ian and Will, Sebastian},
  journal = {PRX Quantum},
  volume = {6},
  issue = {2},
  pages = {020358},
  numpages = {25},
  year = {2025},
  month = {Jun},
  publisher = {American Physical Society},
  doi = {10.1103/b8pm-3prn},
  url = {https://link.aps.org/doi/10.1103/b8pm-3prn}
}

@article{Tang2018,
  title = {Tuning the Dipole-Dipole Interaction in a Quantum Gas with a Rotating Magnetic Field},
  author = {Tang, Yijun and Kao, Wil and Li, Kuan-Yu and Lev, Benjamin L.},
  journal = {Phys. Rev. Lett.},
  volume = {120},
  issue = {23},
  pages = {230401},
  numpages = {5},
  year = {2018},
  month = {Jun},
  publisher = {American Physical Society},
  doi = {10.1103/PhysRevLett.120.230401},
  url = {https://link.aps.org/doi/10.1103/PhysRevLett.120.230401}
}

@article{Giovanazzi2002,
  title = {Tuning the Dipolar Interaction in Quantum Gases},
  author = {Giovanazzi, Stefano and G\"orlitz, Axel and Pfau, Tilman},
  journal = {Phys. Rev. Lett.},
  volume = {89},
  issue = {13},
  pages = {130401},
  numpages = {4},
  year = {2002},
  month = {Sep},
  publisher = {American Physical Society},
  doi = {10.1103/PhysRevLett.89.130401},
  url = {https://link.aps.org/doi/10.1103/PhysRevLett.89.130401}
}

@misc{SM,
note={See the Supplemental Material (at the url provided by the publisher) for details on the perturbation theory.}
}

@article{Balents2010,
  author    = {Balents, Leon},
  title     = {Spin liquids in frustrated magnets},
  journal   = {Nature},
  volume    = {464},
  pages     = {199--208},
  year      = {2010},
  doi       = {10.1038/nature08917}
}

@article{Hikihara2010,
  title = {Magnetic phase diagram of the spin-$\frac{1}{2}$ antiferromagnetic zigzag ladder},
  author = {Hikihara, Toshiya and Momoi, Tsutomu and Furusaki, Akira and Kawamura, Hikaru},
  journal = {Phys. Rev. B},
  volume = {81},
  issue = {22},
  pages = {224433},
  numpages = {20},
  year = {2010},
  month = {Jun},
  publisher = {American Physical Society},
  doi = {10.1103/PhysRevB.81.224433},
  url = {https://link.aps.org/doi/10.1103/PhysRevB.81.224433}
}

@article{Dasgupta2026,
  title = {Chiral phases and dynamics of dipoles in triangular optical ladders},
  author = {Dasgupta, Arjo and \L{}\k{a}cki, Mateusz and Korbmacher, Henning and Dom\'{\i}nguez-Castro, Gustavo A. and Zakrzewski, Jakub and Santos, Luis},
  journal = {Phys. Rev. A},
  volume = {113},
  issue = {3},
  pages = {L031301},
  numpages = {6},
  year = {2026},
  month = {Mar},
  publisher = {American Physical Society},
  doi = {10.1103/wp7z-tg3v},
  url = {https://link.aps.org/doi/10.1103/wp7z-tg3v}
}

@article{Majumdar1970,
doi = {10.1088/0022-3719/3/4/019},
url = {https://doi.org/10.1088/0022-3719/3/4/019},
year = {1970},
month = {apr},
publisher = {},
volume = {3},
number = {4},
pages = {911},
author = {C K Majumdar},
title = {Antiferromagnetic model with known ground state},
journal = {Journal of Physics C: Solid State Physics}
}

@article{Batista2009,
  title = {Canted spiral: An exact ground state of $XXZ$ zigzag spin ladders},
  author = {Batista, C. D.},
  journal = {Phys. Rev. B},
  volume = {80},
  issue = {18},
  pages = {180406},
  numpages = {4},
  year = {2009},
  month = {Nov},
  publisher = {American Physical Society},
  doi = {10.1103/PhysRevB.80.180406},
  url = {https://link.aps.org/doi/10.1103/PhysRevB.80.180406}
}

@article{Batista2012,
  title = {Condensation of Anyons in Frustrated Quantum Magnets},
  author = {Batista, C. D. and Somma, Rolando D.},
  journal = {Phys. Rev. Lett.},
  volume = {109},
  issue = {22},
  pages = {227203},
  numpages = {5},
  year = {2012},
  month = {Nov},
  publisher = {American Physical Society},
  doi = {10.1103/PhysRevLett.109.227203},
  url = {https://link.aps.org/doi/10.1103/PhysRevLett.109.227203}
}

@article{Baier2016,
author = {S. Baier  and M. J. Mark  and D. Petter  and K. Aikawa  and L. Chomaz  and Z. Cai  and M. Baranov  and P. Zoller  and F. Ferlaino },
title = {Extended Bose-Hubbard models with ultracold magnetic atoms},
journal = {Science},
volume = {352},
number = {6282},
pages = {201-205},
year = {2016},
doi = {10.1126/science.aac9812},
URL = {https://www.science.org/doi/abs/10.1126/science.aac9812}}

@article{Su2023,
	Author = {Su, Lin and Douglas, Alexander and Szurek, Michal and Groth, Robin and Ozturk, S. Furkan and Krahn, Aaron and H{\'e}bert, Anne H. and Phelps, Gregory A. and Ebadi, Sepehr and Dickerson, Susannah and Ferlaino, Francesca and Markovi{\'c}, Ognjen and Greiner, Markus},
	Da = {2023/10/01},
	Doi = {10.1038/s41586-023-06614-3},
	Id = {Su2023},
	Isbn = {1476-4687},
	Journal = {Nature},
	Number = {7984},
	Pages = {724--729},
	Title = {Dipolar quantum solids emerging in a Hubbard quantum simulator},
	Ty = {JOUR},
	Url = {https://doi.org/10.1038/s41586-023-06614-3},
	Volume = {622},
	Year = {2023}}

@Article{Hauschild2018,
	title={{Efficient numerical simulations with Tensor Networks: Tensor Network Python (TeNPy)}},
	author={Johannes Hauschild and Frank Pollmann},
	journal={SciPost Phys. Lect. Notes},
	pages={5},
	year={2018},
	publisher={SciPost},
	doi={10.21468/SciPostPhysLectNotes.5},
	url={https://scipost.org/10.21468/SciPostPhysLectNotes.5},
}

@article{Morera2023,
  title = {Superexchange Liquefaction of Strongly Correlated Lattice Dipolar Bosons},
  author = {Morera, Ivan and O\l{}dziejewski, Rafa\l{} and Astrakharchik, Grigori E. and Juli\'a-D\'{\i}az, Bruno},
  journal = {Phys. Rev. Lett.},
  volume = {130},
  issue = {2},
  pages = {023602},
  numpages = {6},
  year = {2023},
  month = {Jan},
  publisher = {American Physical Society},
  doi = {10.1103/PhysRevLett.130.023602},
  url = {https://link.aps.org/doi/10.1103/PhysRevLett.130.023602}
}
\setcounter{equation}{0}
\setcounter{figure}{0}
\renewcommand{\theequation}{S\arabic{equation}}
\renewcommand{\thefigure}{S\arabic{figure}}

\clearpage

\onecolumngrid

\section{Supplemental Material}

In this supplemental material we provide further details concerning anyon condensates, the perturbation theory when departing from the exactly-sovable case, and the engineering of the proper conditions for the observation of anyon condensates using itinerant dipolar bosons in triangular optical ladders.

\section{Perturbation Theory}

The $J_1$-$J_2$ Hamiltonian of Eq.(1) can be written as 
$\hat H_{J_1-J_2}=\sum_j \hat H_j$, where 
\begin{eqnarray}
\hat H_j = E_+ \hat\Pi_{+,j}  + E_- \hat\Pi_{-,j} + E_A \hat\Pi_{A,j} ,
\label{eq:Hj}
\end{eqnarray}
is defined on the triangular plaquette $P_j \equiv \{j-1,j,j+1\}$, and 
$\hat \Pi_{\beta=\{+,-,A\},j} 
= |\xi_{\beta,j}^\uparrow\rangle\langle \xi_{\beta,j}^\uparrow| 
+ |\xi_{\beta,j}^\downarrow\rangle\langle \xi_{\beta,j}^\downarrow|$
are projectors to the one-magnon states:
\begin{eqnarray}
\!\!\!\!\!\!\!\!|\xi_{+,j}^\downarrow\rangle \!&=&\! \cos\alpha\! \left ( \frac{|\uparrow,\downarrow,\downarrow\rangle \!+\! |\downarrow,\downarrow,\uparrow\rangle}{\sqrt{2}}\right ) \!+\!\sin\alpha |\downarrow,\uparrow,\downarrow\rangle, \\
\!\!\!\!\!\!\!\!|\xi_{-,j}^\downarrow\rangle \!&=&\! -\sin\alpha\! \left ( \frac{|\uparrow,\downarrow,\downarrow\rangle \!+\! |\downarrow,\downarrow,\uparrow\rangle}{\sqrt{2}}\right ) \!+\!\cos\alpha |\downarrow,\uparrow,\downarrow\rangle, \\
\!\!\!\!\!\!\!\!|\xi_{A,j}^\downarrow\rangle \!&=&\!\frac{|\uparrow,\downarrow,\downarrow\rangle \!-\! |\downarrow,\downarrow,\uparrow\rangle}{\sqrt{2}},
\end{eqnarray}
and the two-magnon states:
\begin{eqnarray}
\!\!\!\!\!\!\!\!|\xi_{+,j}^\uparrow\rangle \!&=&\! \cos\alpha \!\left ( \frac{|\downarrow,\uparrow,\uparrow\rangle \!+\! |\uparrow,\uparrow,\downarrow\rangle}{\sqrt{2}}\right ) \!+\!\sin\alpha |\uparrow,\downarrow,\uparrow\rangle, \\
\!\!\!\!\!\!\!\!|\xi_{-,j}^\uparrow\rangle \!&=&\! -\sin\alpha \!\left ( \frac{|\downarrow,\uparrow,\uparrow\rangle \!+\! |\uparrow,\uparrow,\downarrow\rangle}{\sqrt{2}}\right ) \!+\!\cos\alpha |\uparrow,\downarrow,\uparrow\rangle,  \\
\!\!\!\!\!\!\!\!|\xi_{A,j}^\uparrow\rangle \!&=&\!\frac{|\downarrow,\uparrow,\uparrow\rangle \!-\! |\uparrow,\uparrow,\downarrow\rangle}{\sqrt{2}},
\end{eqnarray}
where $\tan(2\alpha)=\frac{2\Omega}{\lambda_1-\lambda_2}$, with $\lambda_1=\frac{2J_2(1-\Delta_2)-J_1\Delta_1}{4}$, 
$\lambda_2=-\frac{J_1\Delta_1}{2}$, and $\Omega=\frac{J_1}{2\sqrt{2}}$. In Eq.~\eqref{eq:Hj}, 
\begin{eqnarray}
E_\pm &=& \left (\frac{\lambda_1+\lambda_2}{2} \right )\pm 
\sqrt{\left (\frac{\lambda_1-\lambda_2}{2}\right)^2+\Omega^2}, \\
E_A &=& -\left (\frac{2J_2(1+\Delta_2)+J_1\Delta_1}{4}\right ).
\end{eqnarray}
If $J_1=-4J_2\cos Q$ and $\Delta_\nu=\cos\nu Q$, then $E_{-,A}=0$, and 
$\hat H_{J_1-J_2} = E_+  \sum_j \hat\Pi_{+,j}$.  Since $E_+>0$, all states annihilated by $\sum_j \hat\Pi_{+,j}$ are exact ground states of $\hat H_{J_1-J_2}$. One may show that anyon condensates $|\psi_{n,m}(Q)\rangle$ with any $n$ and $m$ are annihilated by the projectors if $\phi=-4Q$. 
Deviating from the exactly-solvable conditions, 
results in $E_{-,A}\neq 0$. For small-enough $|\delta|$ we may however assume 
$|E_{-,A}|<E_+$, and evaluate the first-order energy correction: $E_{n,m}=\sum_j E_{n,m;j}$, with 
$E_{n,m;j} = \sum_{\beta=-,A}  E_\beta \sum_{\sigma=\uparrow,\downarrow} |\langle \xi_{\beta,j}^\sigma | \psi_{n,m}(Q)\rangle|^2$. 

Let us consider first the contribution of states which have only one ($Q$)-magnon in the plaquette $j$, i.e. the magnon is either in $j-1$, $j$ or $j+1$, and there is only one in that plaquette. Concerning the rest of the magnons, there are  
$m$ ($-Q$)-magnons in $L-3$ plaquettes (since there is no ($-Q$) magnon in the plaquette $j$) and 
$n-1$ ($Q$)-magnons in $L-3-m$ sites. There are hence 
$\binom{L-3-m}{n-1}\binom{L-3}{m}= \binom{L-3}{m+n-1}\binom{m+n-1}{m}$ states of the other magnons which fulfill that there is only one ($Q$) magnon in the $j$ plaquette. Then:
\begin{eqnarray}
|\langle \xi_{\beta,j}^\downarrow | \psi_{n,m}(Q)\rangle|^2 = 
\frac{\binom{L-3}{m+n-1}\binom{m+n-1}{m}}
{ \binom{L}{m+n}\binom{m+n}{m}} \left  |\langle \xi_{\beta,j}^\downarrow | \left [ e^{-iQ} |\uparrow,\downarrow,\downarrow\rangle 
+ |\downarrow,\uparrow,\downarrow\rangle + e^{iQ} |\downarrow,\downarrow,\uparrow\rangle \right ] \right |^2.
\end{eqnarray} 
If $L\gg 1$:
\begin{eqnarray}
|\langle \xi_{\beta,j}^\downarrow | \psi_{n,m}(Q)\rangle|^2 \simeq 
\frac{n}{L} (1-\rho)^2\left  |\langle \xi_{\beta,j}^\downarrow | \left [ e^{-iQ} |\uparrow,\downarrow,\downarrow\rangle 
+ |\downarrow,\uparrow,\downarrow\rangle + e^{iQ} |\downarrow,\downarrow,\uparrow\rangle \right ] \right |^2,
\end{eqnarray} 
with $\rho$ the magnon density.
The same correction comes from the ($-Q$)-magnons, but now $m$ instead of $n$ in the prefactor. The energy contribution coming from the case of a plaquette with a single magnon is then:

\begin{eqnarray}
E_{n,m;j}^\downarrow \simeq \rho (1-\rho)^2\left [ E_- \left ( 1+2\cos^2(Q)\right ) + 2 E_A \sin^2(Q)\right ],  
\label{eq:E_down}
\end{eqnarray}

which is a constant if we consider a fixed $\rho$.

Let us consider now the contribution of states which have two magnons in the plaquette $j$. 
Within the anyon-condensate wavefunction, the contribution of states with two magnons in plaquette $j$ is of the form:
$|\psi_{n,m}\rangle_{2M} = |\psi_{n,m}\rangle_{Q,Q}+|\psi_{n,m}\rangle_{-Q,-Q}+ |\psi_{n,m}\rangle_{Q,-Q}$. 
The states with two ($Q$)-magnons in plaquette $j$ are of the form:
\begin{eqnarray}
|\psi_{n,m}\rangle_{QQ} \propto |\psi_{Q,Q}\rangle \otimes \sum_{\mathbf{j}} e^{i\varphi(\mathbf{j})} | \mathbf{j}\rangle
\end{eqnarray}
with $|\psi_{Q,Q}\rangle = e^{2iQ} \left [e^{-iQ} |\uparrow,\uparrow,\downarrow\rangle +  
|\uparrow,\downarrow,\uparrow\rangle + e^{iQ} |\downarrow,\uparrow,\uparrow\rangle \right ]$, 
and the sum goes over all $\binom{L-3}{m}\binom{L-3-m}{n-2}$ Fock states $|\mathbf{j}\rangle$ with $(n-2)$ ($Q$)-magnons and $m$ ($-Q$)-magnons in the rest of $(L-3)$ sites.

The states with two ($-Q$)-magnons in plaquette $j$ are of the form:
\begin{eqnarray}
|\psi_{n,m}\rangle_{-Q,-Q} \propto |\psi_{-Q,-Q}\rangle \otimes \sum_{\mathbf{j}} e^{i\varphi'(\mathbf{j})} | \mathbf{j}\rangle
\end{eqnarray}
with $|\psi_{-Q,-Q}\rangle = e^{-2iQ} \left [e^{iQ} |\uparrow,\uparrow,\downarrow\rangle +  
|\uparrow,\downarrow,\uparrow\rangle + e^{-iQ} |\downarrow,\uparrow,\uparrow\rangle \right ]$, 
and the sum goes over all $\binom{L-3}{n}\binom{L-3-n}{m-2}$ Fock states $|\mathbf{j}\rangle$ with $n$ ($Q$)-magnons and $(m-2)$ ($-Q$)-magnons. 

The states with one ($Q$)-magnon and one ($-Q$) magnon in plaquette $j$ are of the form:
\begin{eqnarray}
|\psi_{n,m}\rangle_{Q,-Q} \propto |\psi_{Q,-Q}\rangle \otimes \sum_{\mathbf{j}} e^{i\varphi''(\mathbf{j})} | \mathbf{j}\rangle
\end{eqnarray}
with $|\psi_{Q,-Q}\rangle = e^{-2iQ} \left [\left (e^{-iQ}+e^{iQ}e^{i\phi}\right ) |\uparrow,\uparrow,\downarrow\rangle +  
\left (e^{-2iQ}+e^{2iQ}e^{i\phi}\right )|\uparrow,\downarrow,\uparrow\rangle + \left (e^{-iQ}+e^{iQ}e^{i\phi}\right ) |\downarrow,\uparrow,\uparrow\rangle \right ]$, 
and the sum goes over all $\binom{L-3}{n-1}\binom{L-3-n}{m-1}$ Fock states $|\mathbf{j}\rangle$ with $(n-1)$ ($Q$)-magnons and $(m-1)$ ($-Q$)-magnons. 

We need to trace over the states of all other sites but those of the plaquette $j$. The reduced density matrix would be then
$\rho_{2M} = \sum_{\mathbf{n}}\langle \mathbf{n}|\psi_{n,m}\rangle_{2M} \langle\psi_{n,m}|_{2M} | \mathbf{n} \rangle$, 
where we sum over all spin states of the spins outside the plaquette. Then:
\begin{eqnarray}
\!\!\!\!\!\!\!\rho_{2M} \!=\! \frac{\binom{L-3}{m+n-2}}{\binom{L}{m}\binom{L-m}{n}} \!\left [
\binom{m\!+\!n\!-\!2}{m}|\psi_{Q,Q}\rangle \langle \psi_{Q,Q}|
\!+\!\binom{m\!+\!n\!-\!2}{m-2}|\psi_{-Q,-Q}\rangle \langle \psi_{-Q,-Q}|
\!+\!\binom{m\!+\!n\!-\!2}{m\!-\!1}|\psi_{Q,-Q}\rangle \langle \psi_{Q,-Q}|
\right ]
\end{eqnarray} 
Note that in principle there may be cross terms of the form e.g. 
$\propto |\psi_{Q,Q}\rangle \langle \psi_{-Q,-Q}| \sum_{\mathbf{j}} \sum_{\mathbf{j}'} \delta_{\mathbf{j},\mathbf{j}'} e^{i(\varphi(\mathbf{j})-\varphi'(\mathbf{j}'))}$. But the phase factors are to a good approximation quasi-random, and hence 
they average to zero. As a result the cross terms cancel. 
Considering $L\gg 1$, we have then:
\begin{eqnarray}
\rho_{2M} \simeq  (1-\rho) \left [ \frac{n^2}{L^2} |\psi_{Q,Q}\rangle \langle \psi_{Q,Q}|
+\frac{m^2}{L^2} |\psi_{-Q,-Q}\rangle \langle \psi_{-Q,-Q}|
+\frac{mn}{L^2}|\psi_{Q,-Q}\rangle \langle \psi_{Q,-Q}|
\right ].
\end{eqnarray} 
We may then evaluate 
\begin{eqnarray}
|\langle \xi_{-,j}^\uparrow | \psi_{n,m}\rangle |^2 &\simeq&  (1-\rho) 
\left [ \frac{n^2}{L^2} |\langle \xi_{-,j}^\uparrow | \psi_{Q,Q}\rangle|^2
+\frac{m^2}{L^2} |\langle \xi_{-,j}^\uparrow | \psi_{-Q,-Q}\rangle|^2
+\frac{mn}{L^2} |\langle \xi_{-,j}^\uparrow | \psi_{Q,-Q}\rangle|^2
\right ] \nonumber \\
&=& (1-\rho) (1+2\cos^2Q)\left ( \rho^2 + \frac{2nm}{L^2}\right )
\end{eqnarray} 

Proceeding similarly we obtain:
\begin{eqnarray}
|\langle \xi_{A,j}^\uparrow | \psi_{n,m}\rangle |^2 &\simeq& 
2(1-\rho)\sin^2Q\left (\rho^2-\frac{2nm}{L^2} \right )
\end{eqnarray}

Then, summing the contribution of one- and two-magnon states, we get for $L\gg 1$, the energy per plaquette:
\begin{eqnarray}
E_j(\rho,n,m) = \rho(1-\rho) R_+(Q, \Delta_1, \Delta_2) + 2\rho^2 (1-\rho)R_-(Q, \Delta_1, \Delta_2) \frac{nm}{L^2},  
\end{eqnarray}
with
\begin{eqnarray}
R_\pm(Q, \Delta_1, \Delta_2) &=& E_- \left (1+2\cos^2(Q) \right ) \pm 2 E_A \sin^2(Q),
\end{eqnarray} 

If we consider a fixed density, the first term is a constant and we just need to consider 
$E_{j}(n,m) \propto R_-(Q, \Delta_1, \Delta_2) \frac{nm}{L^2}$. 
Then, if $R_-(Q, \Delta_1, \Delta_2)>0$, the energy is minimized by taking $n=0,m=N$ or $n=N,m=0$, i.e. the CSF. 
But if $R(Q, \Delta_1, \Delta_2)<0$ then the energy minimum occurs when $m=n=N/2$, i.e. a balanced 
anyon condensate.
For the case $\Delta_\nu=(1+\delta)\cos\nu Q$ discussed in the main text, the energy correction that depends on $n$ and $m$ is for $\delta<0$ of the form:
\begin{equation}
\Delta E_{n,m} = J_2 (1-4\cos^2 Q)|\delta|\frac{nm}{L^2},
\end{equation}
as in Eq.~(5) of the main text. For $\delta>0$ the formation of self-bound solutions renders inapplicable the first-order calculation.

\section{Engineering the conditions for anyon condensates}

\begin{figure*}[h!]
\centering
\includegraphics[width=0.9\textwidth]{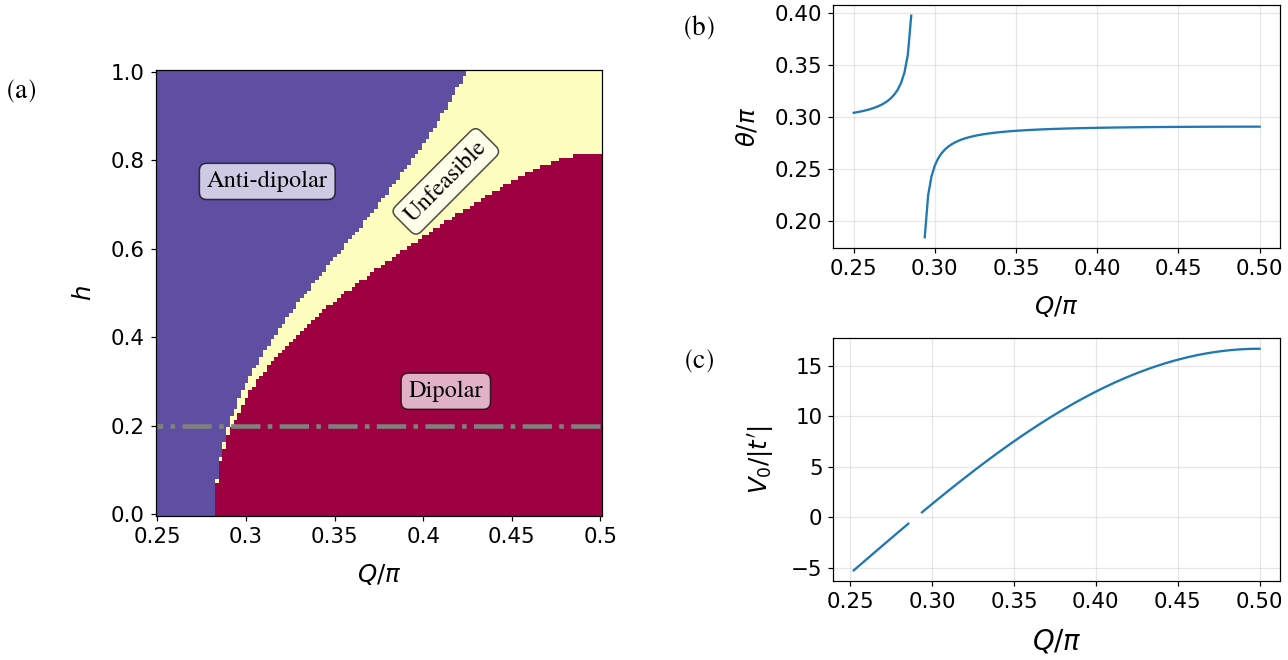}
\caption{(a) Values of $(Q,h)$ requiring dipolar or antidipolar interactions, as well as unfeasible points.
(b) Required $\theta$ as a function of $Q$ for $h=0.2$. 
(c) Required value of dipolar strength $V_0$ as a function of $q$ for $h=0.2$.}
\label{fig:S1}
\end{figure*}

As mentioned in the main text, the exactly-solvable 
conditions may be engineered using hard-core dipoles in triangular ladder by choosing a proper dipole orientation~($\theta$), and strength~($V_0$). 
As discussed in the main text, for a given value of $Q$ and of the 
inter-leg separation $h\sqrt{3}/2$, 
the exactly-solvable model demands $V_0=V_{0,c}(Q,h)$, and $\cos^2\theta_c = F(q,h)$. 
Fig~\ref{fig:S1}~(a) shows the values of $(Q,h)$ requiring dipolar~($V_{0,c}>0$) or antidipolar~($V_{0,c}<0$) configurations, along with unfeasible points~(for which $|F(q,h)|>1$). Examples of curves 
$V_0=V_{0,c}(Q,h)$ and $\theta=\theta_c(Q,h)$
are displayed in Figs.~\ref{fig:S1}~(b) and (c).



\end{document}